\documentclass[12pt]{article}

\usepackage{amsmath,amssymb}
\usepackage{graphicx}

\makeatletter

\renewcommand\section{\@startsection{section}{1}{\z@}
                                   {-3.5ex \@plus -1ex \@minus -.2ex}
                                   {2.3ex \@plus .2ex}
                                   {\normalfont\large\bfseries}}
\renewcommand\subsection{\@startsection{subsection}{2}{\z@}
                                   {-3.25ex\@plus -1ex \@minus -.2ex}
                                   {1.5ex \@plus .2ex}
                                   {\normalfont\normalsize\bfseries}}
\renewcommand\subsubsection{\@startsection{subsubsection}{3}{\z@}
                                   {-3.25ex\@plus -1ex \@minus -.2ex}
                                   {1.5ex \@plus .2ex}
                                   {\normalfont\normalsize\bfseries}}
\renewcommand\paragraph{\@startsection{paragraph}{4}{\z@}
                                   {3.25ex \@plus1ex \@minus.2ex}
                                   {-1em}
                                   {\normalfont\normalsize\bfseries}}

\makeatother

\newcommand{\be}{\begin{equation}}
\newcommand{\ee}{\end{equation}}
\newcommand{\bea}{\begin{eqnarray}}
\newcommand{\eea}{\end{eqnarray}}
\newcommand{\ba}{\begin{array}}
\newcommand{\ea}{\end{array}}

\newcommand{\id}{\hbox{1\kern-.27em l}}
\newcommand{\ZZ}{\mathbb{Z}}
\newcommand{\RR}{\mathbb{R}}

\newcommand{\al}{\alpha}
\newcommand{\ga}{\gamma}
\newcommand{\Ga}{\Gamma}

\newcommand{\cN}{\mathcal{N}}

\newcommand{\N}{M}

\newcommand{\non}{\nonumber}

\newcommand{\SU}{\mathrm{SU}}
\newcommand{\SO}{\mathrm{SO}}
\newcommand{\SL}{\mathrm{SL}}
\newcommand{\Sp}{\mathrm{Sp}}

\newcommand{\so}{\mathrm{so}}
\newcommand{\spl}{\mathrm{sp}}
\newcommand{\Spin}{\mathrm{Spin}}

\newcommand{\ts}{\textstyle}

\newcommand{\beq}{\begin{equation}}
\newcommand{\eeq}{\end{equation}}

\begin{document}

\begin{center}

\vspace*{5mm}
{\Large\sf  Zero-energy states of $\cN = 4$ SYM on $T^3$\,: \\ 
$S$-duality and the mapping class group}

\vspace*{5mm}
{\large M{\aa}ns Henningson, Niclas Wyllard}

\vspace*{5mm}
Department of Fundamental Physics\\
Chalmers University of Technology\\
S-412 96 G\"oteborg, Sweden\\[3mm]
{\tt mans,wyllard@fy.chalmers.se}     
     
\vspace*{5mm}{\bf Abstract:} 
\end{center}

\noindent
We continue our studies of the low-energy spectrum of $\cN=4$ super-Yang-Mills theory on a spatial three-torus. In two previous papers, we computed the spectrum of normalizable zero-energy states for all choices of gauge group and all values of the electric and magnetic 't Hooft fluxes, and checked its invariance under the $\SL_2 (\ZZ)$ $S$-duality group. In this paper, we refine the analysis by also decomposing the space of bound states into irreducible unitary representations of the $\SL_3 (\ZZ)$ mapping class group of the three-torus. We perform a detailed study of the $S$-dual pairs of theories with gauge groups $\Spin(2n{+}1)$ and $\Sp(2n)$. The predictions of $S$-duality (which commutes with the mapping class group) are fulfilled as expected, but the proof requires some surprisingly intricate combinatorial infinite product identities.

\setcounter{equation}{0}
\section{Introduction}

The $\cN = 4$ supersymmetric Yang-Mills theory has no mass gap; the spectrum of the theory extends continuously down to zero energy. This is true not only in Minkowski space, but also if the theory is considered on $\RR \times T^3$,  where the first factor denotes time and the second is a spatial three-torus. In the weak coupling limit, the wave functions of the low-energy states are then supported on the moduli space of flat connections on the gauge bundle over $T^3$. (This means that the magnetic field strength is zero.) Generically, these low-energy states break the gauge group $G$ of the theory to an abelian subgroup, but on certain subspaces of the moduli space of flat connections, the unbroken subgroup $Z$ may be of the form
\beq
Z \simeq S \times U (1)^r ,
\eeq
where $S$ is a semi-simple group and $r$ is some non-negative integer. Because of the scalar fields in the  $\cN = 4$ multiplets associated with the abelian $U(1)^r$ factor, the quantum states will in general not be normalizable; we refer to this as a rank $r$ continuum of states. The effective low-energy theory associated with the semi-simple factor $S$ is modeled by supersymmetric quantum mechanics with 16 supercharges based on the Lie algebra $s$ of the group $S$ \cite{Witten:2000}. 
The latter theory is the dimensional reduction to $0+1$ dimensions of $\cN = 4$ Yang-Mills theory, and is believed to have a linear space $V_s$ of normalizable zero-energy states \cite{Witten:1995}. In the Yang-Mills theory, there are thus $\sum_s \dim V_s$ continua of rank $r$ of low-energy states. These can be further characterized by their discrete abelian 't Hooft fluxes \cite{'tHooft:1979}: The magnetic 't Hooft flux $m$ measures the topological class of the gauge bundle, and the electric 't Hooft flux $e$ determines (together with the $\theta$-angle) the transformation properties of the quantum states under large gauge transformations. 

In two previous papers \cite{Henningson:2007a,Henningson:2007b}, we constructed the low-energy spectrum of normalizable states for all choices of a simple gauge group $G$ (assuming the spectrum is independent of the coupling constant). In particular, we showed that the requirement of $SL_2 (\ZZ)$ $S$-duality   \cite{Goddard:1976}   (under which the 't Hooft fluxes $(m, e)$ transform as a doublet) determines the dimensions of the spaces $V_s$ for all semi-simple Lie algebras $s$ (almost) uniquely:  $\dim V_s$ equals the number of distinguished markings of the corresponding Dynkin diagram. 
(This is in agreement with the result obtained by considering a mass-deformed version of the $\cN=4$ quantum mechanics and assuming that the states are independent of the mass perturbation~\cite{Kac:1999b}.)

In this paper, we refine the analysis of the spectrum by also examining the behavior of the states under the $\SL_3 (\ZZ)$ mapping class group of the spatial $T^3$: For given values of the 't Hooft fluxes $m$ and $e$, the corresponding space of quantum states may be decomposed as a direct sum of irreducible unitary representations $R$ of the stability subgroup of the mapping class group that leaves $m$ and $e$ invariant. The spectrum of degeneracies of such representations should be invariant under $S$-duality acting on $m$ and $e$, but this is not at all manifest in the present formulation of the theory. 
We believe that it should eventually be possible to give a simpler proof of this invariance under $S$-duality, valid for continua of states of arbitrary rank $r$ and all gauge groups $G$. Such a result is likely to give additional insight into the structure of the theory. Here, we proceed in a more pedestrian way, though, and limit ourselves to truly normalizable zero-energy states, i.e.~the $r = 0$ case, and the $S$-dual pair $G = \Spin (2 n {+}1)$ and $G = \Sp (2 n)$. (The case $G = \SU (n)$ is rather trivial \cite{Henningson:2007a}: For given values of $m$ and $e$, there is at most one state, and this transforms trivially under the stability subgroup of the mapping class group. We also have some partial results for the $G = \Spin (2 n)$ cases, but as they shed no particular further light on the underlying structure, we have chosen not to present them here. The cases when $G$ is an exceptional Lie group appear to be technically complicated, but should otherwise pose no particular problems. However, it would probably be more worthwhile to try to understand the general structures, rather than proceeding in a case-by-case manner.)

After a short description of the general aspects of the theory in section two, we compute the spectrum of bound states of the $\Spin (2 n {+} 1)$ and $\Sp (2 n)$ theories in sections three and four respectively. The results are then compared in section five, and found to agree with the predictions of $S$-duality. The proof reveals surprising connections to subtle combinatorial identities related to infinite product expressions for theta functions. The deeper meaning of this is still unclear to us.

\setcounter{equation}{0}
\section{General considerations}
Let $G$ be the gauge group with center subgroup $C$. 
By choosing a specific basis of three primitive one-cycles on $T^3$, the isomorphism class of a principal $G / C$ bundle over $T^3$ (the gauge bundle) may be specified by a triple
\be
m = (m_{23}, m_{31}, m_{12}) \in C^3 ,
\ee
where $m_{ij} = m_{ji}^{-1} \in C$ is identified with the restriction of the discrete abelian magnetic 't Hooft flux to a two-torus in the $ij$-plane. It is sometimes convenient to use the dual notation
\be
m = (m_1, m_2, m_3) \in C^3.
\ee
A flat connection on such a bundle is determined by its holonomies along the non-trivial cycles of the torus, i.e.~by a triple
\be
U = (U_1, U_2, U_3) \in G^3 ,
\ee
subject to the almost commutation relations
\be
U_i U_j U_i^{-1} U_j^{-1} = m_{ij} .
\ee
Let $[U]$ denote the equivalence class of $U$ modulo simultaneous (gauge) conjugation of the $U_i$ by some element of $G$. For a detailed description of the structure of the moduli space of flat connections, see \cite{Borel:1999}.

For an almost commuting triple $U$, we let $Z \subset G$ denote its centralizer (the unbroken gauge group), i.e.~the subgroup of elements that commute with the $U_i$. For simplicity, we will in this paper only be concerned with $U$ such that $Z = S$ is semi-simple, i.e.~we will not consider continua of non-zero rank $r$. For such a $U$, we let $\Delta$ be the finite set of distinguished markings of the Dynkin diagram associated to the Lie algebra $s$ of $S$.  As described in the introduction, there is a linear space $V_s$ of normalizable zero-energy states in the supersymmetric quantum mechanics with 16 supercharges based on $s$. $V_s$ has an orthonormal basis in one-to-one correspondence with the elements of $\Delta$.  For a fixed isomorphism class of semi-simple centralizer $Z$, we then get a linear space $V_Z$ of normalizable zero-energy states with an orthonormal basis of elements denoted $\left| [U], \delta \right>$.  Here $[U]$ is a conjugacy class of an almost commuting triple with semi-simple centralizer isomorphic to $Z$, and $\delta \in \Delta$. The total space $V$ of bound states is the direct sum of the spaces $V_Z$, where the sum runs over all possible semi-simple centralizers $Z \subset G$.

The $C^3$ group of $G / C$ gauge transformations with a non-trivial winding around the cycles of $T^3$ is a module of the $\SL_3 ( \mathbb Z)$ mapping class group of $T^3$, so we may form the semi-direct product 
\be
\hat{\Omega} = \SL_3 ( \mathbb Z) \ltimes   C^3 . \label{Omega-hat}
\ee
This group (almost) acts by permutations on the set of $U$, and this action preserves the centralizer $Z$: The action of the first factor (the mapping class group) is induced from the action on the homology of $T^3$, so that the group element
\be
 A = \left(\ba{ccc} a_{11} & a_{12} & a_{13} \\ a_{21} & a_{22} & a_{23} \\ a_{31} & a_{32} & a_{33} \ea \right) \in SL_3 (\ZZ)
\ee
acts according to
\beq
\left(
\begin{matrix}
U_1 \cr 
U_2 \cr 
U_3 
\end{matrix}
\right)
\mapsto
\left(
\begin{matrix}
U_1^{a_{11}} U_2^{a_{12}} U_3^{a_{13}} \cr 
U_1^{a_{21}} U_2^{a_{22}} U_3^{a_{23}} \cr
U_1^{a_{31}} U_2^{a_{32}} U_3^{a_{33}} 
\end{matrix}
\right) .
\eeq
(This is well-defined if $U$ is a commuting triple, but for an almost commuting triple we need to consider $U$ modulo conjugation by elements of the finite group generated by the $U_i$. Such conjugations act trivially on the centralizer $Z$.) This implies that the components of $m$ transform as a triplet under $\SL_3 (\ZZ)$. The action of the second factor in (\ref{Omega-hat}) is
\beq
\left(
\begin{matrix}
U_1 \cr 
U_2 \cr 
U_3 
\end{matrix}
\right)
\mapsto
\left(
\begin{matrix}
c_1 U_1 \cr 
c_2 U_2 \cr
c_3 U_3 
\end{matrix}
\right) ,
\eeq
where $c_i \in C$. Both factors act trivially on $\delta$. This action on $U$ and $\delta$ descends to an action on pairs $([U], \delta)$, which however may be non-trivial also on $\delta$. Indeed, a choice of preferred representatives $U$ for the classes $[U]$ is in general not preserved by the action of $\hat{\Omega}$, so each such transformation must be accompanied by a suitable conjugation in $G$. This defines an automorphism of $Z$, which, modulo conjugation in $Z$, determines a Dynkin diagram automorphism acting on $\delta$. 

The action of $\hat{\Omega}$ on the pairs  $([U], \delta)$ induces a linear action on the vector space $V$ of normalizable zero-energy states. We begin the analysis of this action by considering the factor $C^3$ in $\hat{\Omega}$. An irreducible linear representation of $C^3$ is determined by a triple (the discrete abelian electric 't Hooft flux)
\be
e = (e_1, e_2, e_3) \in \tilde{C}^3 \simeq C^3 ,
\ee
transforming in the same way as $m$ under $\SL_3 ( \mathbb Z)$. Here we have used the (in this context canonical) isomorphism between the finite abelian group $C$ and its dual $\tilde{C} = {\rm Hom} (C, U (1))$. We may thus decompose the space $V$ of normalizable zero-energy states as a direct sum of subspaces, each of which is characterized by an orbit of $\SL_3 (\mathbb Z)$ on the set of ordered pairs $(m, e) \in C^3 \times C^3$, together with an irreducible representation $R$ of the `little' subgroup $\Omega_{m, e} \subset \SL_3 (\mathbb Z)$ stabilizing some chosen pair on that orbit. We let $N_{m, e}^R (G)$ denote the multiplicity of such representations. The $\SL_2 (\mathbb Z)$ $S$-duality group is expected to commute with the $\SL_3 (\mathbb Z)$ mapping class group and transforms the pair $(m, e)$ as a doublet. Invariance of the spectrum under $S$-duality thus amounts to the conditions
\bea
N_{c, c^\prime}^R (G) & = & N_{c^\prime, c^{-1}}^R (G^\prime) \cr
N_{c, c^\prime}^R (G) & = & N_{c, c c^\prime}^R (G) ,
\eea
where $G^\prime$ denotes the Langlands or GNO dual group of $G / C$. It is not obvious that the spectrum fulfills these conditions, but in the following sections we will check this explicitly for the $S$-dual pairs of theories with gauge groups $\Spin (2 n  + 1)$ and $\Sp (2 n)$.

It should be noted that $N_{m, e}^R (G)$ necessarily vanishes for certain combinations of $m$ and $e$: For a given $m \in C^3$, multiplication of a triple $U = (U_1, U_2, U_3)$ by $(m_{1i}, m_{2i}, m_{3i})$ for some $i = 1, 2, 3$ is equivalent to conjugation by $U_i$, and thus acts trivially on the space of states. The possible values of $e \in \tilde{C}^3$ are thus those that obey $e_1 (m_{1i}) e_2 (m_{2i}) e_3 (m_{3i}) = 1 \in U (1)$. (This notation means that the components of $e$ are evaluated on the components of $m$.) Evaluating this equation for $i = 1, 2, 3$, and using the dual notation for $m$, gives three equations that can be summarized as 
\be
e_i m_j = e_j m_i \label{em-constraint}
\ee
for all $i, j = 1, 2, 3$. The multiplicity $N_{m, e}^R (G)$ can thus be non-zero only for $m$ and $e$ fulfilling this $S$-duality covariant constraint.

It should also be noted that, although the infinite discrete group $\hat{\Omega}$ has infinitely many inequivalent representations, only finitely many of these will appear. The reason is that the holonomies $U_i$ with semi-simple centralizers are of finite order in $G$ (i.e. a finite power of $U_i$ equals the identity element). There is therefore a normal subgroup $\hat{\Omega}_0$ of finite index in $\hat{\Omega}$ that acts trivially on the holonomies, so all representations of $\hat{\Omega}$ that appear are pullbacks of representations of the finite quotient group
\beq
\Gamma = \hat{\Omega} / \hat{\Omega}_0 .
\eeq
Representation theory of finite groups will therefore play a central role in our analysis; for some background material on such groups see e.g.~\cite{Hamermesh:1989}. To decompose the space $V$ of normalizable zero-energy states as a direct sum of irreducible representations of $\Gamma$, we can proceed as follows: For $\gamma \in \Gamma$ we let $[\gamma]$ denote its conjugacy class, i.e.~the set of elements obtained from $\ga$ by conjugation by elements of the group. The cardinality of $[\gamma]$ is denoted $d_{[\gamma]}$, so that 
\beq
d = \sum_{[\gamma]} d_{[\gamma]} ,
\eeq
where the sum runs over all conjugacy classes, equals the order of $\Gamma$. For each irreducible unitary representation $R$ of $\Gamma$ and each conjugacy class $[\gamma]$, the character of $R$ evaluated on $[\gamma]$ is given by ${\rm Tr}_R (\gamma)$, where $\gamma$ is a representative of $[\gamma]$. The multiplicity $N^R$ of the representation $R$ in the decomposition of $V$ now follows from the orthogonality properties of the characters, and is given by the formula
\beq
N^R = \frac{1}{d} \sum_{[\gamma]} d_{[\gamma]} {\rm Tr}_V (\gamma) {\rm Tr}_R (\gamma) , \label{orthogonality}
\eeq
where the sum again runs over all conjugacy classes. It follows from the structure of $V$ as described above, that the trace ${\rm Tr}_V (\gamma)$ equals the number of  pairs $([U], \delta)$ fixed by the action of $\gamma$. This means that $[U]$ should be invariant under $\gamma$, and that $\delta$ should be invariant under the Dynkin diagram automorphism induced by $\gamma$ as described above.

\setcounter{equation}{0}
\section{The $G = \Spin (2 n + 1)$ theories}
In this section we discuss the $G=\Spin(2n{+}1)$ theories; see~\cite{Henningson:2007a} for relevant background material.
As discussed in the introduction, the bound states arise at points in the moduli space of flat connections where the unbroken gauge group is semi-simple. Possible such semi-simple centralizers are of the form~\cite{Witten:1998,Keurentjes:2000,Henningson:2007a}
\beq
z \simeq \so (k_0) \oplus \so (k_1) \oplus \ldots \oplus \so (k_7) \,,
\eeq
where $k_0 + k_1 + \ldots k_7 = 2 n + 1$. One may think of $k_1,  
\ldots, k_7$ as being associated with the points of the Fano plane,  
i.e.~the non-zero points of $\mathbb Z_2^3$ (which can be viewed as the corners of a cube). When $m \in C^3 \simeq  
\mathbb Z_2^3$ is trivial, there are two possibilities~\cite{Witten:1998}: Either $k_0$  
is even and the $k_a$ for $a = 1, \ldots, 7$ are odd, or vice versa.  
When $m \in C^3 \simeq \mathbb Z_2^3$ takes one of the seven 
non-trivial values, it determines one of the seven lines of the Fano  
plane. Again there are two possibilities: Either the $k_a$ associated  
to the three points of that line are even whereas $k_0$ and the  
remaining four $k_a$ are odd, or vice versa. In all cases, the images  
of the corresponding commuting triples in $G / C \simeq \SO (2 n + 1) 
$ may be represented by  the diagonal matrices 
\bea
\bar{U}_1  &=& \mathrm{diag}(\id_{k_0}, -\id_{k_1},  \id_{k_2} , -\id_{k_3},  \id_{k_4}, -\id_{k_5} , \id_{k_6}, -\id_{k_7})  \non \\
\bar{U}_2 &=& \mathrm{diag}(\id_{k_0}, \id_{k_1},  -\id_{k_2} , -\id_{k_3},  \id_{k_4}, \id_{k_5} , -\id_{k_6}, -\id_{k_7}) \\
\bar{U}_3 &=& \mathrm{diag}(\id_{k_0}, \id_{k_1}, \id_{k_2} ,  \id_{k_3} ,-\id_{k_4}, -\id_{k_5}, -\id_{k_6}, -\id_{k_7}) \non
\eea
The order of the above holonomies $\bar{U}_i$ in $G / C$ is $1$ or $2$ so  
that $U_i^2 \in C$. In a sector with a fixed value of $e$, it is  
therefore sufficient to consider the action on the holonomies of a  
finite group $\Gamma$ defined as the reduction modulo 2 of the  
stability group $\Omega_{m, e} \subset \SL_3 (\mathbb Z)$.

There are $2^3 = 8$ different liftings of a triple from $\SO (2 n {+}1) 
$ to $\Spin (2 n {+} 1)$, related by multiplication with elements of  
$C^3$, but generically these define the same holonomy modulo  
conjugation. This means that the corresponding states all have $e$  
trivial. However, if the four $k_a$ associated with the points that  
do not belong to a certain line of the Fano plane are zero, then  
multiplication by the corresponding non-trivial element of $C^3$  
changes the equivalence class of the triple. There will then be a  
further set of states with a corresponding non-trivial value 
of $e$ \cite{Henningson:2007a}.

It follows that for $m \in C^3$ trivial, all values of $e \in C^3$  
are possible. A non-trivial value of $e$ only appears when $k_0$ is  
odd, the four $k_a$ associated with points not in the corresponding  
line are zero, and the three $k_a$ associated with points in that  
line are even. For $m \in C^3$ non-trivial, $e$ is either trivial or  
equal to $m$. The latter case appears only when $k_0$ is even, the  
four $k_a$ associated with points not in the corresponding line are  
zero, and the remaining three $k_a$ are odd.  These pairs $(m, e) \in  
C^3 \times C^3$ are precisely those allowed by (\ref{em-constraint}),  
namely
\beq
\begin{array}{ll}
(m,e) & \quad\Gamma \\[3pt]
\hline\\[-9pt] 
(0, 0) & \SL_3 (\mathbb Z_2) \cr
(0, c) & \SL_2 (\mathbb Z_2) \cr
(c, 0) & \SL_2 (\mathbb Z_2)  \ltimes \mathbb Z_2^2 \cr
(c, c) & \SL_2 (\mathbb Z_2) ,
\end{array}
\eeq
where $0$ denotes the trivial element of $\mathbb Z_2^3$ and $c$ is  
one of the seven non-trivial elements. We have also indicated the  
relevant finite quotient $\Gamma$ of the stability group $\Omega_{m,  
e}$. It acts on the holonomies by permuting the $k_a$. (For $(m, e) = (0, c)$ one might
think that the relevant group is $\SL_2 (\mathbb Z_2)  \ltimes \mathbb Z_2^2$, 
but the second factor acts trivially on the holonomies.)

It is advantageous to consider all values of the rank of the gauge group, $n$, simultaneously: The  
number of distinguished markings of the $\so (k)$ Dynkin diagram,  
i.e. $\dim V_{\so (k)}$, equals the number of partitions of $k$ into  
distinct odd parts \cite{Kac:1999b},  and this is most easily described by the  
generating function
\beq
P (q) = \sum_{k = 1}^\infty q^k \dim V_{\so (k)} = \prod_{k = 1}^ 
\infty (1 + q^{2 k - 1}) \,,
\eeq
which we decompose into its even and odd powers $P_{\rm even} = \frac 
{1}{2} (P (q) + P (-q))$ and $P_{\rm odd} = \frac{1}{2} (P (q) - P (- 
q))$, respectively.
So rather than directly determining the multiplicities of bound states transforming in the unitary representation $R$ of $\Ga$, denoted by $\N_{m, e}^R$ in this section, we will work with the generating functions
\beq
\N_{m, e}^R (q) = \sum_{n = 1}^\infty q^{2 n + 1} \N_{m, e}^R \,.
\eeq
These functions can be computed by a formula analogous to (\ref{orthogonality}):
\beq
\N_{m, e}^R (q) = \frac{1}{d} \sum_{[\gamma]} d_{[\gamma]} T_{m, e}^ 
{[\gamma]} (q) {\rm Tr}_R (\gamma) \,, \label{TN-conversion}
\eeq
where
\beq
T_{m, e}^{[\gamma]} (q) = \sum_{n = 0}^\infty q^{2 n + 1} {\rm Tr}_V  
(\gamma)\, .
\eeq
In the latter formula, the trace is over the space $V$ of  
normalizable zero-energy states in the $(m, e)$-sector of the $\Spin  
(2 n + 1)$ theory.

We will now carry out these computations for all possible values of $m 
$ and $e$ separately.

\subsection{The $(m, e) = (0, 0)$ states}
The group $\Gamma = \SL_3 (\mathbb Z_2)$ is of order $d = 168$. The  
values of ${\rm Tr}_R (\gamma)$ are given in the character table:
\beq
\begin{array}{crrrrrr}
{\rm \bf conj. class} & {\bf 1^7} & {\bf 1^3 \, 2^2} & {\bf 1 \, 2 \,  
4} & {\bf 1 \, 3^2} & {\bf 7} & {\bf 7^\prime} \cr
{\rm cardinality} & 1 & 21 & 42 & 56 & 24 & 24 \\[3pt]
\hline \\[-3pt]
{\bf 1} & 1 & 1 & 1 & 1 & 1 & 1 \cr
{\bf 6} & 6 & 2 & 0 & 0 & -1 & -1 \cr
{\bf 7} & 7 & -1 & -1 & 1 & 0 & 0 \cr
{\bf 8} & 8 & 0 & 0 & -1 & 1 & 1 \cr
{\bf 3} & 3 & -1 & 1 & 0 & \bar{c} & c  \cr
{\bf \bar{3}} & 3 & -1 & 1 & 0 & c & \bar{c}
\end{array}
\eeq
where $c = \frac{1}{2} \left(-1 - i \sqrt{7}\right)$ and $\bar{c}$ is its complex conjugate. We have denoted the representations by their dimensionality in  bold face, and  
the conjugacy classes by their cycle structure when acting on the seven $k_a$.  
 From the above considerations follows that
\bea
T_{0, 0}^{\bf 1^7} (q) & = & P_{\rm odd} (q) P_{\rm even}^7 (q) + P_{\rm even}  
(q) P_{\rm odd}^7 (q) \cr
T_{0, 0}^{\bf 1^3  2^2} (q) & = & P_{\rm odd} (q) P_{\rm even}^3 (q) P_{\rm even} 
^2 (q^2) + P_{\rm even} (q) P_{\rm odd}^3 (q) P_{\rm odd}^2 (q^2) \cr
T_{0, 0}^{\bf 1  2  4} (q) & = & P_{\rm odd} (q) P_{\rm even} (q) P_{\rm even}  
(q^2) P_{\rm even} (q^4) + P_{\rm even} (q) P_{\rm odd} (q) P_{\rm odd} (q^2) P_{\rm odd}  
(q^4) \cr
T_{0, 0}^{\bf 1  3^2} (q) & = & P_{\rm odd} (q) P_{\rm even} (q) P_{\rm even}^2  
(q^3) + P_{\rm even} (q) P_{\rm odd} (q) P_{\rm odd}^2 (q^3) \cr
T_{0, 0}^{\bf 7} (q) & = & P_{\rm odd} (q) P_{\rm even} (q^7) + P_{\rm even} (q)  
P_{\rm odd} (q^7) \cr
T_{0, 0}^{\bf 7^\prime} (q) & = & P_{\rm odd} (q) P_{\rm even} (q^7) + P_ 
{\rm even} (q) P_{\rm odd} (q^7) .
\eea
The first (second) term in each expression corresponds to $k_0$ being  
odd (even) and the $k_a$ being even (odd). The generating functions  
$\N_{0, 0}^R (q)$ are obtained through (\ref{TN-conversion}).
As an example we give the resulting expression for the six-dimensional 
representation (here we have also used the second identity in (\ref{ids})):
\bea
\N^{\bf 6}_{0,0}&=&{\ts \frac{1}{3584}}P(q)^8 
+{\ts \frac{1}{128}} P(q)^4 P(q^2)^2 
+{\ts \frac{3}{256}} P(q)^4 P(-q^2)^2 \non \\
&+&{\ts \frac{1}{32}} P(q)^2 P(-q^2)P(-q^4) 
+{\ts \frac{1}{7}} P(q) P(q^7) -(q\leftrightarrow -q)\,.
\eea
Note that $\sum_R \dim R \N_{0, 0}^R (q)$ reproduces the result in~\cite{Henningson:2007a} as required for consistency.

\subsection{The $(m, e) = (0, c)$ states}
The group $\Gamma = \SL_2 (\mathbb Z_2) \simeq S_3$ (the symmetric  
group on three elements) is of order $d = 6$. Its character table is
\beq
\begin{array}{crrr}
{\rm \bf conj. class} & {\bf 1^3} & {\bf 1 \, 2} & {\bf 3} \cr
{\rm cardinality} & 1 & 3 & 2 \\[3pt]
\hline \\[-3pt]
{\bf \!1}  & 1 & 1 & 1 \cr
{\bf 1^\prime} & 1 & -1 & 1 \cr
{\bf \!2} & 2 & 0 & -1
\end{array} 
\eeq
where we have denoted the conjugacy classes by their cycle structure  
on the three non-zero $k_a$. The $k_a$ are necessarily even, and $k_0$ is  
odd. It follows that
\bea
T^{\bf 1^3}_{0, c} (q) & = & P_{\rm odd} (q) P_{\rm even}^3 (q) \cr
T^{\bf 1 2}_{0, c} (q) & = & P_{\rm odd} (q) P_{\rm even} (q) P_{\rm even} (q^2) \cr
T^{\bf 3}_{0, c} (q) & = & P_{\rm odd} (q) P_{\rm even} (q^3) \,.
\eea
The generating functions $\N_{0, c}^R (q)$ follow from  (\ref{TN-conversion}).

\subsection{The $(m, e) = (c, 0)$ states}
The group $\Gamma = \SL_2 (\mathbb Z_2)  \ltimes \mathbb Z_2^2 \simeq  
S_4$ (the symmetric group on four elements) is of order $d = 24$. Its  
character table is
\beq
\begin{array}{crrrrr}
{\rm \bf conj. class} & {\bf (1^3, 1^4)} & {\bf (1^3,  2^2)} & {\bf  
(1 \, 2, 1^2 \, 2)} & {\bf (1 \, 2, 4)} & {\bf (3, 1 \, 3)} \cr
{\rm cardinality} & 1 & 3 & 6 & 6 & 8 \\[3pt]
\hline \\[-3pt]
{\bf \!1}  & 1 & 1 & 1 & 1 & 1\cr
{\bf 1^\prime} & 1 & 1 & -1 & -1 & 1 \cr
{\bf \!2} & 2 & 2 & 0 & 0 & -1 \cr
{\bf \!3} & 3 & -1 & 1 & -1 & 0 \cr
{\bf 3^\prime} & 3 & -1 & -1 & 1 & 0 
\end{array}
\eeq
The two entries in the notation for the conjugacy classes refer to to  
the cycle structures on the three $k_a$ in the line on the Fano plane  
and the four remaining $k_a$ respectively. We get
\bea
\!T^{\bf (1^3, 1^4)}_{c, 0} (q) & = & P_{\rm odd}^3 (q) P_{\rm even}^5 (q) + P_ 
{\rm even}^3 (q) P_{\rm odd}^5 (q) \cr
\!T^{\bf (1^3,  2^2)}_{c, 0} (q) & = & P_{\rm odd}^3 (q) P_{\rm even} (q) P_ 
{\rm even}^2 (q^2) + P_{\rm even}^3 (q) P_{\rm odd} (q) P_{\rm odd}^2 (q^2) \cr
\!T^{\bf (1 \, 2, 1^2 \, 2)}_{c, 0} (q) & = & P_{\rm odd} (q) P_{\rm odd} (q^2)  
P_{\rm even}^3 (q) P_{\rm even} (q^2) + P_{\rm even} (q) P_{\rm even} (q^2) P_{\rm odd}^3  
(q) P_{\rm odd} (q^2) \cr
\!T^{\bf (1 \, 2, 4)}_{c, 0} (q) & = & P_{\rm odd} (q) P_{\rm odd} (q^2) P_ 
{\rm even} (q) P_{\rm even} (q^4) + P_{\rm even} (q) P_{\rm even} (q^2) P_{\rm odd} (q) P_ 
{\rm odd} (q^4) \cr
\!T^{\bf (3, 1 \, 3)}_{c, 0} (q) & = & P_{\rm odd} (q^3) P_{\rm even}^2 (q) P_ 
{\rm even} (q^3) + P_{\rm even} (q^3) P_{\rm odd}^2 (q) P_{\rm odd} (q^3) ,
\eea
where the first (second) term in each expression corresponds to the  
three $k_a$ in the line on the Fano plane being odd (even) and the  
four remaining $k_a$ together with $k_0$ being even (odd). The  
generating functions $\N_{c, 0}^R (q)$ follow from  (\ref{TN-conversion}).

\subsection{The $(m, e) = (c, c)$ states}
The group $\Gamma = \SL_2 (\mathbb Z_2) \simeq S_3$ is the same as in the $(m, e) = (0, c)$ case, but  
now the three non-zero $k_a$ are odd, and $k_0$ is even. It follows that
\bea
T^{\bf 1^3}_{c, c} (q) & = & P_{\rm even} (q) P_{\rm odd}^3 (q) \cr
T^{\bf 1 2}_{c, c} (q) & = & P_{\rm even} (q) P_{\rm odd} (q) P_{\rm odd} (q^2) \cr
T^{\bf 3}_{c, c} (q) & = & P_{\rm even} (q) P_{\rm odd} (q^3) .
\eea
The generating functions $\N_{c, c}^R (q)$ follow from  (\ref{TN-conversion}).

\setcounter{equation}{0}
\section{The $G=\Sp(2n)$ theories}
In this section we perform an analysis of the $\Sp(2n)$ theories, similar to the one carried out for the $\Spin(2n{+}1)$ theories in the previous section (see~\cite{Henningson:2007a} for relevant background material). It turns out that it is convenient to treat all $e$ values together; we therefore split the analysis into two cases, $m=0$ and $m\neq 0$.
\subsection{The $m=0$ states}
The bound states with $m=0$ arise from semi-simple centralizers of the form~\cite{Witten:2000,Henningson:2007a}
\be
z \simeq \spl(2k_1) \oplus \spl(2k_2) \oplus \cdots \oplus \spl(2k_8)\,,
\ee
where $\sum_i k_i = n$.
One may think of $k_1,\ldots,k_8$ as being associated with the eight corners of a cube, or equivalently $\ZZ_2^3$ (see \cite{Henningson:2007a} for more details).   
The corresponding commuting triples may be represented by the diagonal matrices \cite{Henningson:2007a}
\bea \label{Sppair}
U_1  &=& \mathrm{diag}(\id_{2k_1}, -\id_{2k_2},  \id_{2k_3} , -\id_{2k_4},  \id_{2k_5}, -\id_{2k_6} , \id_{2k_7}, -\id_{2k_8}) \,, \non \\
U_2 &=& \mathrm{diag}(\id_{2k_1}, \id_{2k_2},  -\id_{2k_3} , -\id_{2k_4},  \id_{2k_5}, \id_{2k_6} , -\id_{2k_7}, -\id_{2k_8})\,, \\
U_3 &=& \mathrm{diag}(\id_{2k_1}, \id_{2k_2}, \id_{2k_3} ,  \id_{2k_4} ,-\id_{2k_5}, -\id_{2k_6}, -\id_{2k_7}, -\id_{2k_8}) \,. \non
\eea
Since $U_i^2=1$ the action of $\SL(3,\ZZ)$ is reduced to $\SL(3,\ZZ_2)$. The large gauge transformations act as reflections in the cube language and induce the natural $SL(3,\ZZ_2)$ action on $e = (e_1,e_2,e_3)$. These facts imply that the natural group to use to classify the $m=0$ states is $\SL(3,\ZZ_2)\ltimes\ZZ_2^3$. This group has order 1344 and its character table is\footnote{This table is derived in~\cite{Littlewood:2006}, or can be obtained using the GAP computational algebra system~\cite{GAP4}.}
\be
\begin{array}{crrrrrrrrrrr}
{\rm \bf conj. class} & {\bf 1^8} & {\bf 1^42^2}  & {\bf 2^4} & {\bf 2^{4\prime}} & {\bf 1^2 3^2} &{\bf 1^224} & {\bf 26} & {\bf 17}& {\bf 17^\prime} &{\bf 4^2} & {\bf 4^{2^\prime}}\cr
{\rm cardinality} &1 & 42 & 42 & 7 & 224 & 168 & 224 & 192 & 192 & 168 & 84 \\[3pt]
\hline \\[-3pt]
{\bf 1} & 1 & 1 & 1 & 1 & 1 & 1 & 1 & 1 & 1 & 1 & 1 \\
{\bf 3}& 3 & -1 & -1 & 3 & 0 & 1 & 0 & c  & \bar{c}  & 1 & -1 \\
{\bf \bar{3}}& 3 & -1 & -1 & 3 & 0 & 1 & 0 & \bar{c}  & c  & 1 & -1 \\
{\bf 6}&  6 & 2 & 2 & 6 & 0 & 0 & 0 & -1 & -1 & 0 & 2 \\
{\bf 7} &7 & -1 & -1 & 7 & 1 & -1 & 1 & 0 & 0 & -1 & -1 \\
{\bf 8}&  8 & 0 & 0 & 8 & -1 & 0 & -1 & 1 & 1 & 0 & 0 \\
{\bf 1} \cdot 7& 7 & 3 & -1 & -1 & 1 & 1 & -1 & 0 & 0 & -1 & -1 \\
{\bf 1^\prime} \cdot 7 &7 & -1 & 3 & -1 & 1 & -1 & -1 & 0 & 0 & 1 & -1 \\
{\bf 2} \cdot 7 & 14 & 2 & 2 & -2 & -1 & 0 & 1 & 0 & 0 & 0 & -2 \\
{\bf 3} \cdot 7& 21 & 1 & -3 & -3 & 0 & -1 & 0 & 0 & 0 & 1 & 1 \\
{\bf 3^\prime} \cdot 7 &21 & -3 & 1 & -3 & 0 & 1 & 0 & 0 & 0 & -1 & 1 
\end{array}
\ee
where $c = \frac{1}{2} \left(-1 - i \sqrt{7}\right)$ and $\bar{c}$ is its complex conjugate. The non-trivial large gauge transformations constitute the conjugacy class ${\bf 2^{4^\prime}}$ of cardinality $7$. We see that these act trivially, i.e.~imply $e = 0$, on the first six representations, which therefore can be identified with representations of the $\SL_3 (\ZZ_2)$ stability group. The last five representations contain the states where $e$ takes one of the 7 non-zero values. We denote these representations as ${\bf R} \cdot 7$, where ${\bf R}$ is a representation of the $\SL_2 (\ZZ_2) \ltimes \mathbb \ZZ_2^2 \simeq S_4$ stability group.

Just as for $\Spin(2n{+}1)$, it is convenient to consider all values of $n$ simultaneously. Indeed, the number of distinguished markings of the $\spl (2 k)$ Dynkin diagram, i.e. $\dim V_{\spl (2k)}$, equals the number of partitions of $2 k$ into distinct even parts. This is most easily described by the generating function
\be
Q (q)  =  \sum_{k = 1}^\infty q^{2 k} \dim V_{\spl (2 k)} = \prod_{n = 1}^\infty (1 + q^{2 n}) \,.
\ee
We can then associate a generating function to each conjugacy class:
\bea \label{Spm0rules}
T_{m=0}^{\bf 1^8} &=& q \,Q(q)^8 \,,\cr
T_{m=0}^{\bf 1^42^2} &=& q\, Q(q)^4Q(q^2)^2 \,,  \cr
T_{m=0}^{\bf 2^4} &=& q \,Q(q^2)^4 \,,\cr
T_{m=0}^{\bf 2^{4\prime}} &=& q\, Q(q^2)^4 \,,\cr
T_{m=0}^{\bf 1^23^2} &=& q \,Q(q)^2 Q(q^3)^2 \,,\cr
T_{m=0}^{\bf 1^224} &=& q \,Q(q)^2Q(q^2)Q(q^4) \,,\cr
T_{m=0}^{\bf 26} &=& q \,Q(q^2)Q(q^6) \,,\cr
T_{m=0}^{\bf 17} &=& q \,Q(q)Q(q^7) \,,\cr
T_{m=0}^{\bf 17^\prime} &=& q\, Q(q)Q(q^7) \,,\cr
T_{m=0}^{\bf 4^2} &=& q \,Q(q^4)^2 \,,\cr
T_{m=0}^{\bf 4^{2\prime}} &=& q \,Q(q^4)^2 \,,
\eea
where the prefactor $q$ is introduced to facilitate the comparison with the $\Spin(2n{+}1)$ results of the previous section. The generating functions for the number of states transforming in the various unitary representations can be computed by a  formula analogous to (\ref{TN-conversion}):
\be \label{method}
N_{m=0}^R (q) = \frac{1}{d} \sum_{[\gamma]} d_{[\gamma]} T_{m=0}^{[\gamma]} (q) {\rm Tr}_R (\gamma) \,. 
\ee
As an example, we write explicitly the generating function for the $(e,m)=(0,0)$ states transforming in the ${\bf 6}$ of $\SL_3 (\ZZ_2)$:
\be
N_{0, 0}^{\bf 6} (q)  =  \frac{q}{224} \left(Q(q)^8{+}14 Q(q)^4 Q(q^2)^2{+}21 Q(q^2)^4{+}28 Q(q^4)^2{-}64 Q(q) Q(q^7)\right)\,.
\ee

\subsection{The $m\neq 0$ states}
The bound states with $m \neq 0$ arise from semi-simple centralizers of 
the form~\cite{Keurentjes:2000,Henningson:2007a}
\be
Z \simeq \so(k_1) \oplus \so(k_1') \oplus \spl(2k_2) \oplus \spl(2k_2') \oplus \cdots \oplus  \spl(2k_4) \oplus \spl(2k_4')\,.
\ee
One may think of the $k_i$ and the $k_i^\prime$ as being associated with the points of two parallel planes (determined by $m$) on a cube. The corresponding holonomies can be found in ~\cite{Henningson:2007a,Keurentjes:2000}. They are at most of order 2, so just as for $m = 0$, only the mod 2 reduction of the $\SL_3 (\ZZ)$ mapping class group is relevant. Furthermore the $\ZZ_2^2$ factor of the $\SL_2 (\ZZ_2) \ltimes Z_2^2$ stability subgroup of $m$ acts trivially on the holonomies, so we need only consider the group $\SL_2 (\ZZ_2)$. Including also the non-trivial large gauge transformations, we are led to consider the finite group 
\beq
\Gamma = \SL_2 (\ZZ_2){\times}\ZZ_2\,,
\eeq
of order $d = 12$. Its character table is
\beq 
\begin{array}{crrrrrr}
{\rm \bf conj. class} & {\bf (1^2, 1^6)} & {\bf (1^2 , 1^2 2^2)} & {\bf (1^2, 3^2)}  & {\bf (2 , 2^3)} & {\bf (2, 2^3)^\prime} & {\bf (2, 6)} \cr
{\rm cardinality} & 1 & 3 & 2 &1 &3 &2 \\[3pt]
\hline \\[-3pt]
{\bf 1_0} & 1 & 1 & 1 &1 &1 &1\cr
{\bf 1^\prime_0} & 1 & -1 & 1 & 1 & -1 & 1 \cr
{\bf 2_0} & 2 & 0 & -1 & 2 & 0 & -1 \cr
{\bf 1_m} & 1 & 1 & 1 & -1 & -1 & -1 \cr
{\bf 1^\prime_m} & 1 & -1 & 1 & -1& 1 & -1 \cr
{\bf 2_m} & 2 & 0 & -1 & -2 & 0 & 1 
\end{array}
\eeq
where the first (second) entry in the notation for the conjugacy classes indicates the cycle structure on $k_1$ and $k_1^\prime$ ($k_2$, $k_3$, $k_4$ and $k_2^\prime$, $k_3^\prime$, $k_4^\prime$). The non-trivial large gauge transformation constitutes the ${\bf (2, 2^3)}$ conjugacy class of cardinality $1$. It is trivially represented on the first three representations, which thus have $e = 0$, and non-trivially represented on the last three representations, which thus have $e = m$.
The generating functions associated with the conjugacy classes are 
\bea
T_{m\neq 0}^{\bf (1^2, 1^6)} &= & q P(q^2)^2 Q(q^2)^6 \,,\non \\
T_{m\neq 0}^{\bf (1^2, 1^2 2^2)} &=&  q P(q^2)^2 Q(q^2)^2 Q(q^4)^2 \,,\non \\
T_{m\neq 0}^{\bf (1^2, 3^2)} &=&  q P(q^2)^2  Q(q^6)^2 \,,\\
T_{m\neq 0}^{\bf (2, 2^3)} &=&  q P(q^4) Q(q^4)^3 \,,\non \\
T_{m\neq 0}^{\bf (2, 2^3)^\prime} &=&  q P(q^4) Q(q^4)^3 \,,\non \\
T_{m\neq 0}^{\bf (2, 6)} &=&  q P(q^4) Q(q^{12}) \non\,.
\eea
The generating functions for the unitary representations are obtained as in (\ref{method}). 

\setcounter{equation}{0}
\section{$S$-duality}
The requirements of $S$-duality are easy to verify by expanding out the expressions obtained in the previous sections for $M_{m,e}^R (q)$ and $N_{m,e}^R (q)$ as power series in $q$. One finds that
\beq
\begin{array}{rr}
M_{0,0}^{\bf 1} = N_{0,0}^{\bf 1} = & q + q^3 + 2 q^5 + 4 q^7 + 8 q^9 + 13 q^{11} + 26 q^{13} + 44 q^{15} + 80 q^{17} + \ldots \cr
M_{0,0}^{\bf 3} = N_{0,0}^{\bf 3} =& 2 q^{15} + 7 q^{17} + \ldots \cr
M_{0,0}^{\bf \bar{3}} = N_{0,0}^{\bf \bar{3}} = & 2 q^{15} + 7 q^{17} + \ldots \cr
M_{0,0}^{\bf 6} = N_{0,0}^{\bf 6} = & q^5 + 2 q^7 + 7 q^9 + 14 q^{11} + 34 q^{13} + 68 q^{15} + 146 q^{17} + \ldots \cr
M_{0,0}^{\bf 7} = N_{0,0}^{\bf 7} = & q^{11} + 6 q^{13} + 16 q^{15} + 43 q^{17} + \ldots \cr
M_{0,0}^{\bf 8} = N_{0,0}^{\bf 8} = & q^9 + 4 q^{11} + 12 q^{13} + 32 q^{15} + 80 q^{17} + \ldots \cr
\cr
M_{0,c}^{\bf 1} = N_{c,0}^{\bf 1} = & q + q^3 + 2 q^5 + 3 q^7 + 7 q^9 + 10 q^{11} + 19 q^{13} + 29 q^{15} + 50 q^{17} + \ldots \cr
M_{0,c}^{\bf 1^\prime} = N_{c,0}^{\bf 1^\prime} = & q^{11} + 3 q^{13} + 7 q^{15} + 13 q^{17} + \ldots \cr
M_{0,c}^{\bf 2} = N_{c,0}^{\bf 2} = & q^5 + 2 q^7 + 5 q^9 + 9 q^{11} + 18 q^{13} + 31 q^{15} + 57 q^{17} + \ldots \cr
\cr
M_{c,0}^{\bf 1} = N_{0,c}^{\bf 1} = & q^3 + 2 q^5 + 6 q^7 + 12 q^9 + 27 q^{11} + 55 q^{13} + 112 q^{15} + 215 q^{17} + \ldots \cr
M_{c,0}^{\bf 1^\prime} = N_{0,c}^{\bf 1^\prime} = & q^9 + q^{11} + 7 q^{13} + 16 q^{15} + 47 q^{17} + \ldots \cr
M_{c,0}^{\bf 2} = N_{0,c}^{\bf 2} = & q^5 + 2 q^7 + 7 q^9 + 18 q^{11} + 45 q^{13} + 100 q^{15} + 222 q^{17} + \ldots \cr
M_{c,0}^{\bf 3} = N_{0,c}^{\bf 3} = & 2 q^7 + 6 q^9 + 19 q^{11} + 46 q^{13} + 116 q^{15} + 257 q^{17} + \ldots \cr
M_{c,0}^{\bf 3^\prime} = N_{0,c}^{\bf 3^\prime} = & q^9 + 5 q^{11} + 18 q^{13} + 52 q^{15} + 137 q^{17} + \ldots \cr
\cr
M_{c,c}^{\bf 1} = N_{c,c}^{\bf 1} = & q^3 + q^5 + 3 q^7 + 5 q^9 + 10 q^{11} + 16 q^{13} + 29 q^{15} + 45 q^{17} + \ldots \cr
M_{c,c}^{\bf 1^\prime} = N_{c,c}^{\bf 1^\prime} = & q^9 + q^{11} + 4 q^{13} + 7 q^{15} + 15 q^{17} + \ldots \cr
M_{c,c}^{\bf 2} = N_{c,c}^{\bf 2} = & q^5 + 2 q^7 + 4 q^9 + 9 q^{11} + 17 q^{13} + 31 q^{15} + 55 q^{17} + \ldots ,
\end{array} \non
\eeq
where $c$ is an arbitrary non-trivial element of $\ZZ_2^3$. This means e.g.~that for each non-trivial $c \in\ZZ_2^3$, the $m = 0$, $e = c$ states of the $\Spin (17)$ theory comprise $57$ doublets of the $\SL_2 (\ZZ_2)$ stability subgroup of the mapping class group.

It is actually possible to prove these identities to all orders in $q$; the analysis reveals a rich structure of combinatorial infinite product identities dating back to Euler, Ramanujan and others.
$S$-duality between the $m=0$ states in the $\Sp(2n)$ theory and the $e =0$ states in the $\Spin(2n+1)$ theory is equivalent to linear combinations of the following identities:
\bea \label{Spe1}
 16 q Q(q)^8 &=& P(q)^8 -  (q\leftrightarrow -q) \,,\non \\
 8 q Q(q)^4 Q(q^2)^2 &=& P(q)^4 P(q^2)^2 - (q\leftrightarrow -q)\,, \non \\
4 q Q(q)^2 Q(q^3)^2 &=& P(q)^2 P(q^3)^2- (q\leftrightarrow -q) \,,\\
4 q Q(q)^2 Q(q^2)Q(q^4) &=& P(q)^2 P(q^2)P(q^4) - (q\leftrightarrow -q) \,,\non \\
2 q Q(q)Q(q^7) &=& P(q)P(q^7) -  (q\leftrightarrow -q) \,,\non
\eea
as well as
\bea \label{Spe2}
8 q Q(q^2)^4 &=& P(q)^4 P(-q^2)^2 -  (q\leftrightarrow -q)\,, \non \\
4 q Q(q^2) Q(q^6) &=& P(q)^2 P(-q^3)^2 - (q\leftrightarrow -q) \,,\\ 
 4 q Q(q^4)^2  &=& P(q)^2 P(-q^2) P(-q^4) - (q\leftrightarrow -q) \,. \non
\eea
Before we proceed let us make a few remarks about these identities. The first identity in (\ref{Spe1}) is Jacobi's famous {\it aequatio identica satis abstrusa}, which also appeared in our previous paper \cite{Henningson:2007a}. The other identities are ``$\SL_3(\ZZ_2)$ refinements'' of this identity. We note that identities of the type (\ref{Spe1}) have recently enjoyed a renewed interest in the mathematics literature. In particular, in the work of Farkas and Kra \cite{Farkas:2000} identity 1, 3 and 5 of (\ref{Spe1}) above were referred to as `a curious property of'  `eight', `three', and `seven', respectively. In their work, these three identities were treated in a case-by-case manner. The fact that we have found a connection via $\SL_3(\ZZ_2)$ and $S$-duality between these and other identities might be of some interest. 

For $S$-duality between the states with $m \neq 0$ in the $\Sp(2n)$ theory and the $e \neq 0$ states in the $\Spin(2n+1)$ theory one similarly needs the identities 
\bea \label{Spm1}
8 q P(q^2)^2 Q(q^2)^6 &=& P(q)^4 -  (q\leftrightarrow -q) \,,\non \\
4 q P(q^2)^2 Q(q^2)^2 Q(q^4)^2 &=& P(q)^2 P(q^2) - (q\leftrightarrow -q) \,, \\
2 q P(q^2)^2 Q(q^6)^2 &=& P(q) P(q^3)- (q\leftrightarrow -q) \non\,,
\eea
as well as
\bea  \label{Spm2}
4 q P(q^4) Q(q^4)^3 &=& P(q)^2 P(-q^2) - (q\leftrightarrow -q) \,,\non \\
2 q P(q^4) Q(q^{12}) &=& P(q)P(-q^3) -  (q\leftrightarrow -q) \,.
\eea
Note that the above identities can be rewritten using the elementary relations
\bea \label{ids}
Q(q^2)P(q^2) = Q(q) \,,\non \\
P(-q)P(q) = P(-q^2)\,,
\eea
together with Euler's identity
\bea \label{Euler}
Q(q) P(-q^2) = 1  \,.  
\eea

All of the above identities are special cases of certain identities among theta functions (they can also be proved starting from entries 29 and 30 in chapter 16 of \cite{Berndt:1991}). 
The following identity
\bea \label{id1}
\!&& 2 q \,h(a,b) \prod_{n=1}^\infty (a {+} q^{2n})(\frac{1}{a}{+}q^{2n}) (b {+} q^{2n})(\frac{1}{b}{+}q^{2n})(a b {+} q^{2n})(\frac{1}{a b}{+}q^{2n})(1 {+} q^{2n})^2  \\
\!&=& 
\prod_{n=1}^\infty (a {+} q^{2n-1})(\frac{1}{a}{+}q^{2n-1}) (b {+} q^{2n-1})(\frac{1}{b}{+}q^{2n-1})(a b {+} q^{2n-1})(\frac{1}{a b}{+}q^{2n-1})(1 {+} q^{2n-1})^2 \non \\
\!&-& \prod_{n=1}^\infty (a {-} q^{2n-1})(\frac{1}{a}{-}q^{2n-1}) (b {-} q^{2n-1})(\frac{1}{b}{-}q^{2n-1})(a b {-} q^{2n-1})(\frac{1}{a b}{-}q^{2n-1})(1 {-} q^{2n-1})^2 \,,\non
\eea
where $h(a,b) = (1+a)(1+b)(1+\frac{1}{ab})$, follows from the theta function result
\bea \label{thetaid0}
&&\theta_2(\al,q) \,\theta_2(\beta,q) \, \theta_2(\al+\beta,q) \, \theta_2(0 ,q) =\\
&&\theta_3(\al,q) \,\theta_3(\beta,q) \, \theta_3(\al+\beta,q) \, \theta_3(0 ,q) -
\theta_4(\al,q)\, \theta_4(\beta,q) \, \theta_4(\al+\beta,q) \, \theta_4(0 ,q)\,. \non
\eea
When is the LHS of (\ref{id1}) of  the form $\prod_{n=1}^\infty \prod_i (1 + q^{2n \ell_i} )$ for integers $\ell_i$ with $ \sum_i \ell_i = 8$? It turns out that there are only five solutions to this requirement, namely 
\beq
(a,b) \in \{ (1,1), (i,1), (-e^{i\pi/3},1),(-e^{i\pi/4},i),(-e^{i \pi/7},-e^{9i\pi/7}) \} . 
\eeq
The resulting identities precisely correspond to the above expressions (\ref{Spe1}). It is interesting to note that essentially the same identity (\ref{id1}) was also needed to show the equality between the number of vacuum states in the $\cN=1^*$  mass-deformed $\cN=4$ $\Sp(2n)$ and $\SO(2n)$ theories, as required by $S$-duality \cite{Wyllard:2007}.

The identity 
\be  \label{thetaid}
2 \,\theta_2(\al+\beta,q^2)\theta_2(\al-\beta,q^2) = 
\theta_3(\al,q)\theta_3(\beta,q) -\theta_4(\al,q)\theta_4(\beta,q) \,, 
\ee 
with $\al=\beta$ can be rewritten as
\bea 
&&4 q \frac{(1+a^2)}{a} \prod_{n=1}^\infty (a^2 + q^{4n}) (\frac{1}{a^2} + q^{4n})(1+q^{4n})^2  \\
&=& 
\prod_{n=1}^\infty (\frac{1}{a}+q^{2n-1})^2 (a + q^{2n-1})^2 (1-q^{4n-2})^2 \non 
- (q\leftrightarrow -q)\,.
\eea
For the two parameter choices 
\beq
a \in \{1,e^{i\pi/3} \} \,, 
\eeq
the resulting identities are identical to the first and second entries in (\ref{Spe2}). 

From the identity 
\be \label{thetaid2} 
2 \theta_2(2\al,q^4)= \theta_3(\al,q)-\theta_4(\al,q)\,, 
\ee 
with $\al=0$ one easily  obtains the final identity in (\ref{Spe2}).

Next we consider the following identity obtained from (\ref{thetaid})
\bea
&&2 q\, g(a,b) \prod_{n=1}^\infty (ab + q^{4n})(\frac{1}{ab}+q^{4n}) (\frac{b}{a} + q^{4n})(\frac{a}{b}+q^{4n})(1 + q^{4n})^2(1+q^{4n-2})  \non \\
&=& 
\prod_{n=1}^\infty (a + q^{2n-1})(\frac{1}{a}+q^{2n-1}) (b + q^{2n-1})(\frac{1}{b}+q^{2n-1})  - (q\leftrightarrow -q)
\eea
where $g(a,b) = (1+\frac{1}{ab})(1+\frac{b}{a})$. There are only three choices of the parameters for which we get integer exponents as above, namely 
\beq
(a,b) \in \{(1,1), (1,i), (1,-e^{i\pi/3}) \} . 
\eeq
The resulting identities precisely correspond to the identities in (\ref{Spm1}). 

From the identity (\ref{thetaid2}) one deduces
\bea
&&2 q \frac{(1+a^2)}{a} \prod_{n=1}^\infty (1 + q^{8n-4})(a^2 + q^{8n}) (\frac{1}{a^2} + q^{8n})(1+q^{8n})  \\
&=& 
\prod_{n=1}^\infty (1 - q^{4n-2})(\frac{1}{a}+q^{2n-1}) (a + q^{2n-1}) \non 
- (q\leftrightarrow -q)
\eea
There are two parameter choices that we need, namely
\beq
a \in \{ -1,e^{i\pi/3} \} . 
\eeq
The resulting expressions reproduce the identities in (\ref{Spm2}). 
Note that the theta function identities used here, (\ref{thetaid0}), (\ref{thetaid}) and (\ref{thetaid2}), are parameter deformations of the identities (5.51-53) used in \cite{Henningson:2007a}.
 
\section*{Acknowledgements}

M.H. is supported by grants from the G\"oran Gustafsson foundation and the Swedish Research Council.\\
N.W. is supported by a grant from the Swedish Research Council.

\begingroup\raggedright\endgroup

\end{document}